\documentclass[sigconf]{acmart}
\AtBeginDocument{%
  }

\setcopyright{acmlicensed}
\copyrightyear{2018}
\acmYear{2018}
\acmDOI{XXXXXXX.XXXXXXX}
\acmConference[Conference acronym 'XX]{Make sure to enter the correct
  conference title from your rights confirmation email}{Sep.,
  2026}{MN}
\acmISBN{978-1-4503-XXXX-X/2018/06}

\usepackage{tcolorbox}
\tcbuselibrary{breakable} 
\usepackage{fancyvrb}
\usepackage{fvextra}
\begin{document}

\title{AgenticRecTune: Multi-Agent with Self-Evolving Skillhub for Recommendation System Optimization}
\author{Xidong Wu}
\email{xidongwu@google.com}
\affiliation{%
  \institution{Google}
  \city{Mountain View}
  \state{California}
  \country{USA}
}

\author{Yue Zhuan}
\email{yuezhuan@google.com}
\affiliation{%
  \institution{Google}
  \city{Mountain View}
  \state{California}
  \country{USA}
}

\author{Ruoqiao Wei}
\email{ruoqiao@google.com}
\affiliation{%
  \institution{Google}
  \city{Mountain View}
  \state{California}
  \country{USA}
}

\author{Hangxin Chen}
\email{hangxinc@google.com}
\affiliation{%
  \institution{Google}
  \city{Mountain View}
  \state{California}
  \country{USA}
}

\author{Di Bai}
\email{vivianbai@google.com}
\orcid{0009-0009-1813-4844}
\affiliation{%
  \institution{Google}
  \city{Mountain View}
  \state{California}
  \country{USA}
}

\author{Jintao Liu}
\email{liujintao@google.com}
\affiliation{%
  \institution{Google}
  \city{Mountain View}
  \state{California}
  \country{USA}
}

\author{Xinyi Wang}
\email{wxinyi@google.com}
\affiliation{%
  \institution{Google}
  \city{Mountain View}
  \state{California}
  \country{USA}
}

\author{Xue Wang}
\email{xuewang@google.com}
\affiliation{%
  \institution{Google}
  \city{Mountain View}
  \state{California}
  \country{USA}
}

\author{Luoshu Wang}
\email{luoshu@google.com}
\affiliation{%
  \institution{Google}
  \city{Mountain View}
  \state{California}
  \country{USA}
}

\author{Xinwu Cheng}
\email{xinwuc@google.com}
\affiliation{%
  \institution{Google}
  \city{Mountain View}
  \state{California}
  \country{USA}
}









\begin{abstract}
Modern large-scale recommendation systems are typically constructed as multi-stage pipelines, encompassing pre-ranking, ranking, and re-ranking phases. While traditional recommendation research typically focuses on optimizing a specific model, such as improving the pre-ranking model structure or ranking models training algorithm, system-level configurations optimization play a crucial role. These system-level configurations integrate the output from each model head to establish the final score or adjusts scores to meet business requirements beyond engagement metrics. .
Due to the complexity of the system, the configuration optimization is highly important and challenging. Any model modification requires new optimal system-level configurations. But each experimental iteration requires significant tuning effort. Furthermore, models in different stage operates within a distinct context and optimizes for different targets, requiring specialized domain expertise. In addition,
optimization success depends on balancing competing multiple online metrics and alignment with shifting production development objectives.

To address these challenges, we propose AgenticRecTune, an agentic framework comprising five specialized agents, Actor, Critic, Insight, Skill, and Online, designed to manage the end-to-end configuration optimization workflow.
By leveraging the advanced reasoning of Large Language Models (LLMs), specifically Google’s Gemini, AgenticRecTune explore the optimal configuration spaces. 
The Actor Agent proposes multiple candidates and Critic Agent filters out suboptimal proposals. Then Online Agent autonomously prepares A/B tests based on the proposed configurations set from the Critic Agent and captures the subsequencet experimental results.
We also introduce a self-evolving Skillhub, which utilizes a collaboration between the Insight Agent and Skill Agent to summarize the history results, extract underlying mechanics of each task in recommendation system and update skills for next round of optimization. 
This holistic framework streamlines the search for optimal parameters and accelerates model deployment. The effectiveness of AgenticRecTune is demonstrated by multiple successful production launches on Google Discover.

\end{abstract}




 \maketitle

\section{Introduction}

\begin{figure*}[h] 
  \centering
  \includegraphics[width=\linewidth]{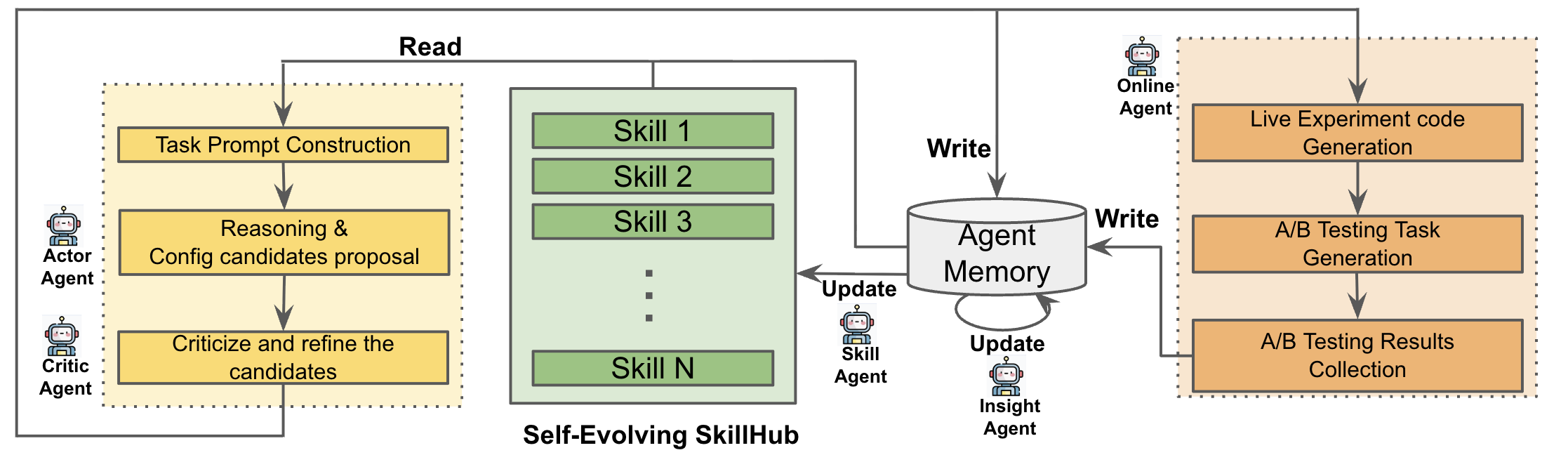}
  \caption{The workflow of AgenticRecTune. The AgenticRecTune optimizes configurations in each component of recommendation system with an agentic architecture (Actor Agent, Critic Agent, Insight Agent, Skill Agent and Online Agent). 
1) AgenticRecTune initiates the process through task prompt construction with reliance on skills and history data from skillhub and agent memory. Then the Actor Agent proposes configuration candidates and the Critic Agent works to criticize and refine the candidates. 2) The Online Agent handles online A/B Testing experiments by managing task generation and test results collection. 3) Data is then Written to the Agent Memory, which is continuously maintained by the Insight Agent and Skill Agent through Updates to the memory and the Self-Evolving SkillHub.
  }
  \label{Workflow}
\end{figure*}
Recommender systems deliver personalized content and are deployed across a wide variety of products, ranging from multi-media content platforms and e-commerce sites to social media networks. Currently, the recommender system is largely formulated as a machine learning (ML) problem, utilizing advanced ML models and ML algorithms to optimize performance. 

To achieve high efficiency and quality at the industrial-scale recommender system, these systems typically employ a multi-stage architecture. Following the initial retrieval phase, a pre-ranking model acts as a fast, lightweight predictor to roughly score and distill a manageable subset of items from the massive candidate pool. Next, the ranking stage applies sophisticated, computationally intensive machine learning models to this reduced subset, precisely ordering items based on predicted user interest. Finally, the re-ranking stage fine-tunes this ordered list before presentation, making final adjustments to ensure content diversity, enforce business logic, and apply final product constraints.

However, although a modern recommender system consists of several interacting components, researchers often isolate and focuses on specific sub-tasks. For instance, researchers frequently treat the ranking stage as an independent multi-task classification problem, designing novel architectures or algorithms specifically to improve engagement metric or long-term metric. 

System-level configurations optimization is crucial for recommender systems. They integrate the outputs from various model heads to derive final rankings at each stage. For instance, ranking models often generate multiple scores per candidate and these are synthesized via complex weighted strategies during value fusion. In this context, the fusion weights serve as the primary system-level configurations that determine overall performance. In addition, in the pre-ranking stage, they also adjusts scores to meet business requirements.
Any structural modification to the underlying models, including a change in feature engineering or a shift in model architecture, invalidates previous settings and necessitates a global recalibration of system-level parameters. 
However, the inherent complexity of these systems makes finding the optimal configurations both vital and highly challenging. Although individual models are optimized using differentiable loss functions, the overarching system is sparse and non-differentiable. Consequently, an alignment gap emerges between the proxy metrics used during training and the system's north star metrics. Specifically, we identify three \textbf{primary challenges} in this optimization process:

\begin{itemize}
\item \textbf{Scalability Limitations}.
Each change in system requires a new set of optimal configurations. The vast optimization space and multi-dimensional dependencies demand expensive grid searches, exhaustive data-driven learning or repeated manual tuning. Consequently, this heavy workflow creates a scalability gap that hinders fast-paced deployment.

\item \textbf{Contextual Fragmentation}. 
The system involves multiple specialized models in different stages, each operating within a distinct context and optimizing for different local targets. Understand the overall system and tuning individual components demand highly specialized domain expertise and substantial engineering experience.

\item \textbf{Challenge of Evolving Multi-Objective Online Metrics}. Model training labels (such as engagement target) primarily focus on isolated targets. 
In practice, however, online success depends on balancing competing metrics (e.g., engagement, diversity, and long-term retention). Furthermore, the system must remain adaptable to evolving product strategies, where the topline metrics are periodically refined to reflect shifting strategic priorities. Directly optimizing for these dynamic, high-dimensional objectives remains a complex task that current paradigms struggle to solve.
\end{itemize}

Overcoming these challenges is beyond the scope of traditional ML methods. Although Automated Machine Learning (AutoML) \cite{zoller2021benchmark} was proposed to enable automatic optimization, standard AutoML techniques \cite{elsken2019neural} are highly effective at optimizing numerical configurations within predefined search spaces, they lack the reasoning skills required to adapt to diverse contexts or directly understand production target with  natural language.

The rapid advancement of Large Language Models (LLMs) has spurred the development of LLM agents designed to tackle these complex ML tasks. 
To solve these problems, we propose AgenticRecTune to facilitates human efforts and improves system performance. AgenticRecTune is deployed across various ranking models within Google Discover, a large-scale multi-media content recommendation platform.
Our primary contributions are summarized as follows:

\begin{itemize}
\item \textbf{End-to-End Configuration Optimization}. 
We propose AgenticRecTune for industrial-Scale recommender system. It is a self-evolving multi-agent framework (including actor agent, critic agent, insight agent, skill agent and online agent) designed to automate the end-to-end system-level configuration optimization workflow with multiple agents. It no longer requires human effort anymore and overcomes the scalability limitations of manually driven iterations.

\item \textbf{Self-Evolving Skillhub} 
We introduce an extensible and self-evolving Skillhub that serves as the agent's repository of domain-specific expertise. Unlike static Skillhub, the Skill Agent continuously ingests learned pattern generated by the Insight Agent from online A/B testing results and long-term system memory. Specifically, the Insight Agent identifies critical parameters and extracts actionable patterns from historical online experiments. The Skill Agent then synthesizes these findings into operational strategies and updates task-specific skills. Through this closed-loop cooperation between the Insight and Skill agents, the system effectively bridges the gap between raw data and executable knowledge, facilitating continuous autonomous self-improvement.

\item \textbf{Optimizing Multiple Online Metrics}. 
AgenticRecTune leverages an Online Agent to operate directly within live experimental environments to navigate the high-dimensional online Pareto front. By bypassing the limitations of offline proxy metrics, the Actor Agent proposes new candidates by dynamically balancing competing objectives, such as engagement, diversity, and long-term retention. This adaptability ensures that the system maintains strict alignment with evolving product development targets.

\item \textbf{Actor-Critic Strategy}. A Critic Agent is introduced to provide verification and filter out suboptimal proposals from the Actor Agent. The collaboration between the Actor Agent and the Critic Agent enhances the stability and convergence of AgenticRecTune, ensuring only the most promising candidates are deployed in live environments.
\end{itemize}

\section{Related Work}
The optimization of industrial-scale recommender systems is currently undergoing a paradigm shift, moving from manually tuned pipelines to autonomous, agent-driven ecosystems as LLM is integrated into recommender systems \cite{wu2024survey}.
We categorize the existing literature into three primary domains to highlight the progression toward autonomous, system-wide optimization.
\subsection{LLMs as Recommenders, Simulators, and Interactive Agents}
Initial efforts focused on leveraging the semantic reasoning of LLMs to improve user preference understanding. The P5 framework \cite{geng2022recommendation} established a unified text-to-text paradigm for diverse tasks, while RecPrompt \cite{liu2024recprompt} automated prompt engineering for news recommendation. To enhance logic, STARec \cite{wu2025starec} introduced "Fast/Slow" cognition for preference reasoning, and MemRec \cite{chen2026memrec} utilized collaborative memory to store user-item pairings. Other frameworks, such as InteRecAgent \cite{huang2025recommender} and RecMind \cite{wang2024recmind}, serve as conversational or planning agents that interact directly with users to refine discovery journeys.

\subsection{Autonomous ML Engineering and Hyperparameter Optimization}
The emergence of autonomous scientific discovery has enabled agents to perform tasks traditionally reserved for Machine Learning Engineers (MLEs). General frameworks like The AI Scientist \cite{lu2024ai} and PACEvolve \cite{yan2026pacevolve} use iterative coding and backtracking to automate research and search execution. In the realm of Hyperparameter Optimization (HPO), AgentHPO \cite{liu2024large} uses a Creator-Executor structure to tune models, while studies on autoresearch \cite{ferreira2026can} show that hybridizing LLMs with classical tracking yields superior results. To improve agent reliability, ML-Agent \cite{liu2025ml} applies reinforcement learning to teach small models how to complete ML engineering tasks.
While effective, these general frameworks primarily optimize for immediate, clear offline rewards (e.g., loss or accuracy) or operate on static natural language constraints like OptimAI \cite{thind2025optimai} and physical control rewards like Eureka \cite{ma2023eureka}. An industrial recommendation pipeline, by contrast, involves multi-tower models and highly non-differentiable thresholds where feedback is noisy and delayed. Standard HPO and general code-editing agents struggle to navigate this operational complexity. Recently, 
\citet{zhang2025agentic} introduce Agentic Context Engineering (ACE), a multi-agent framework that conceptualizes contexts as dynamic playbooks. These playbooks undergo a continuous cycle of generation, reflection, and curation to systematically accumulate and refine strategic knowledge.

\subsection{System-Level Orchestration and Direct Online Metric Optimization}
Recognizing the limitations of general ML agents, recent work has tailored autonomous optimization specifically to recommendation architectures. AgenticTagger \cite{xie2026agentictagger} utilizes an Architect-Annotator loop for automated feature engineering, creating structured item representations. Moving to the model level, Self-EvolveRec \cite{kim2026self} uses diagnosis tools to mutate model source code, while the self-evolving system by \citet{wang2026self} acts as an MLE to rewrite neural architectures and loss functions for YouTube. For objective balancing, DualAgent-Rec \cite{zhang2026llms} uses an LLM coordinator to balance exploitation and exploration within a single stage.
Despite these advancements, existing approaches remain confined to offline preprocessing (feature mining) or model-level structural mutations (editing network layers). Our framework shifts the paradigm from model editing to system-level orchestration and the direct optimization of online metrics. Instead of mutating code or running offline evaluations, our agent acts as an active controller that manages the non-differentiable "glue" of the live system—tuning the fusion weights, routing thresholds, and configurations across the Pre-ranking, Ranking, and Re-ranking stages. Utilizing an extensible Skillhub to manage stage-specific domain knowledge, our system continuously executes live A/B tests to learn from real-world trial metrics. By closing the loop directly with production servers, the agent bypasses proxy offline labels to autonomously maximize global online "North Star" business metrics.

\section{Multi-Level Compositional Optimization Formulation}
Large-scale recommender systems utilize a multi-stage, sequential design that progressively narrows down millions of choices into a highly curated feed. This pipeline typically begins with retrieval stage, which recall a variety of feed for ranking. Then pre-ranking model acts as a fast, lightweight filter to roughly score and select a manageable subset of items—usually in the thousands—from the massive overall pool. Next, the ranking models apply much more sophisticated and computationally intensive machine learning models to this smaller subset, precisely scoring and ordering the items based on their predicted relevance to the user. Finally, the re-ranking stage fine-tunes this ordered list before it is displayed, making final adjustments to ensure content diversity, apply specific business rules, and filter out unwanted or duplicate items. Therefore, large-scale recommender systems includes multiple models and each stage will affect the system performance. 
To achieve better generalization, we define the configurations for each component as below: 
\begin{itemize}
\item Let $\theta_{pre}$ be the configurations of the pre-ranking model.
\item Let $\theta_{rank}$ be the configurations of the ranking model.
\item Let $\theta_{re}$ be the configurations of the re-ranking model.
\end{itemize}

The entire system's configurations space is the joint vector: 
$$\Theta = [\theta_{pre}, \theta_{rank}, \theta_{re}] \in \mathcal{P}$$
Where $\mathcal{P}$ is the system configuration space. Extra configurations can be added if a recommender system is more complex. 

Let $x$ represent the input request (such as user profile, user activities, context features, and the initial massive candidate pool). $\theta_{pre}$,  $\theta_{rank}$ and $\theta_{re}$ denote system configurations in pre-ranking, ranking and re-ranking stage. $\textbf{w}_{pre}$, $\textbf{w}_{rank}$ and $\textbf{w}_{re}$ denote pre-ranking model, ranking model and re-ranking model weights. Each stage in the sequential pipeline acts as a function $f$ that filters or scores the candidates, passing the output to the next stage as the input and is presented as below:
\begin{itemize}
\item \textbf{Pre-Ranking:} $f_{pre}(x; \textbf{w}_{pre}, \theta_{pre})$
\item \textbf{Ranking:} $f_{rank}(f_{pre}, x; \textbf{w}_{rank},\theta_{rank})$
\item \textbf{Re-Ranking:} $ f_{re}(f_{rank}, x; \textbf{w}_{re}, \theta_{re})$
\end{itemize}

The final output of the system (the ranked list of items shown to the user) can be written as a composite function:
$$\mathcal{F} = f_{re}(f_{rank}(f_{pre}(x; \textbf{w}_{pre}, \theta_{pre}), x; \textbf{w}_{rank}, \theta_{rank}), x; \textbf{w}_{re}, \theta_{re})$$

We define $y_{true}$ as the user's actual implicit/explicit actions. Let $M(\mathcal{F} , y_{true})$ be the metric evaluated on the system's output $\mathcal{F}$ against the user's actual implicit/explicit feedback $y_{true}$. The online A/B testing usually includes multiple north star metrics $M = [M_1, ... M_J]$, such as overall DAU, and engagement metrics. We want to maximize the expected values of primary metrics while keep ensuring secondary metrics do not drop below a certain baseline $b_j$, we formulate the optimization target in the recommender system as 
\begin{align}
U(\mathbf{M}) &= \sum_{i=1}^{n} M_1(\mathcal{F}, \mathbf{y}_{true}) \nonumber\\
&\text{s.t.}  M_j(\mathcal{F} , \mathbf{y}_{true}) \ge b_j \quad \forall j \in \{n+1, \dots, J\} \nonumber
\end{align}

Combining all these elements, the tuning of the recommender system is formulated as a multi-task Compositional Optimization \cite{gao2024decentralized, jiang2022optimal} to find the optimal configuration set $\Theta^*$ that maximizes the expected North Star Metrics, subject to system cost $C(\Theta)$ constraints:

\begin{align}
\Theta^* &= \arg\max_{\Theta \in \mathcal{P}} \mathbb{E}_{(x, y_{true}) \sim \mathcal{D}} [U (M(\mathcal{F} (x; \textbf{w}, \Theta), y_{true}))]  \\
&\text{s.t.} \mathbb{E}_{x \sim \mathcal{D}}[C(\Theta)] \le C_{max} \nonumber
\end{align}


\section{AgenticRecTune}
Multi-task Compositional Optimization is notoriously difficult because:  
\begin{itemize}
\item \textbf{Non-Differentiable} The function $F(x; \Theta)$ involves sorting, truncating (Top-K), and business logic. These non-differentiable components prevent the use of standard backpropagation, making it difficult to optimize $\Theta$ through direct gradient-based methods.

\item \textbf{Multi-metric optimization} 
The objective involves simultaneously improving multiple primary metrics while ensuring that secondary guardrail metrics do not degrade. Balancing the trade-offs between these competing objectives presents a significant optimization challenge.
\end{itemize}

To solve this issue, we propose the AgenticRecTune framework to manage and optimize the system configurations based on the reasoning ability of LLM as shown in Figure \ref{Workflow}.

\subsection{Reasoning Loop}
In this section, we detail the iterative process by which agents propose, analyze and refines configuration candidates.

\subsubsection{Task Prompt Construction}
Leveraging the comprehensive context, constraints, and domain knowledge retrieved from the self-evolving skillhub, and elite candidates from Agent Memory, the Actor Agent constructs a highly structured prompt for reasoning and candidate proposal as shown in Prompt \ref{actor_prompt}. To allow for flexible control over the optimization process, users can dynamically inject arguments into this prompt. These extra signals, such as the desired batch size of configuration candidates to generate, or the specific directory for task logging—enable practitioners to adjust the workflow to fit current system constraints.


\begin{tcolorbox}[
    colback=gray!5,
    colframe=gray!50,
    title=Actor Agent Prompt,
    fonttitle=\bfseries,
    breakable
]
\begin{Verbatim}
# ROLE:
You are a Machine Learning Ranking Engineer on the Search Discover. 

# Task CONTEXT : 
1)WHAT IS VALUE-BASED RETRIEVAL?
{task_introduction}

2)Parameters Description: …
# TASK REQUIREMENT:
Propose new configurations to improve topline metrics. 
Analyze the historical data and decide which curve configurations to propose for the next round of live experiment. Find the optimal configuration to meet the evaluation. 

{how_we_evaluate_experiment_metrics}

## North Star Metric:
{multiple_core_business_metrics}
…

## Domain Knowledge:
{domain_knwledge}

{self_learning_pattern}

## Elite Configurations (Pareto Frontier)
Here are the current best-performing (elite) configurations from the Pareto frontier:
{elites_str}

## Initial configuration parameters & Output Format
You should make exactly {max_proposals} new proposals.
<proposals>
  <proposal>
    <hypothesis> {hypothesis_example}</hypothesis>
    <config>
{initial_configuration_example}
    </config>
  </proposal>
</proposals>

# ACTION: propose
Analyze the previous experiments and propose the next batch.
\end{Verbatim}
\label{actor_prompt}
\end{tcolorbox}

\subsubsection{Reasoning and configuration candidates proposal}
AgenticRecTune primarily utilize the Gemini models to power this reasoning. After prompt generation, the actor agent proposes a series of candidate configurations and pays more attention to the sensitive parameters following the instruction. In addition, the actor agent provides a logical explanation for each proposed parameter shift. These explanations ensure that the agent's exploration is interpretable and grounded in the task target, domain knowledge and historical experimentation data provided in the prompt.

\subsubsection{Criticize and refine the candidates}
To ensure the safety and viability of the proposed parameters, a separate evaluates the actor's outputs. The Critic Agent systematically criticizes the proposed candidates from the previous step, checking them against format requirement, system guardrails, instruction constraints, and known historical failure cases as shown in Prompt \ref{refine_prompt}. It filters out suboptimal configurations, selects the most promising candidates for deployment, and provides detailed feedback comments. At the same time, comments are also asked to provide to guarantee the agent follow the instruction correctly. Finally, the Critic Agent writes the candidates to Agent Memory as shown in the Figure \ref{Workflow}.
\begin{tcolorbox}[
    colback=gray!5,
    colframe=gray!50,
    title=Critic Agent Prompt,
    fonttitle=\bfseries,
    breakable
]
\begin{Verbatim}
You are an expert evaluator for Discover feed ranking experiments. Your task is to critique and refine the 'original response' and select the *best {max_proposals}* configurations to move forward with.

The original prompt was:
<original_prompt>
{original_prompt}
</original_prompt>

The original response was:
<original_response>
{original_response}
</original_response>

**Refinement Instructions:**
1.  **Validity Checks:**
   * Ensure NO TYPOS and match the original prompt's intent.
2.  **Alignment with Goals:** Critically evaluate each of the proposals against the "Optimization Objectives" and metric priorities mentioned in the `{original_prompt}`.
3.  **Assess Explanation:** Ensure the `<explanation>` for each config change is logically sound and plausible based on the parameter descriptions.
4.  **Selection Criteria for the Top {max_proposals}:** Choose a diverse set of the most promising proposals. Avoid selecting tuning strategies that are too similar to one another.
5.  **Output Format:**
   *   The output must be a single XML block enclosed in <proposals></proposals> tags.
   *   Output EXACTLY {max_proposals} proposals.
   *   Each selected proposal should be enclosed in `<proposal></proposal>` tags, maintaining the original inner structure (`<hypothesis>` and `<config>`).
   *   **Crucially, add a NEW `<justification>` tag within each selected `<proposal>` block. In this tag, briefly explain WHY this specific proposal was chosen, referencing its potential impact on metrics and the strength of its hypothesis.**

Refined response:
\end{Verbatim}
\label{refine_prompt}
\end{tcolorbox}

\subsection{Online Experiments}
Once the reasoning loop successfully proposes and refines a set of configuration candidates, AgenticRecTune transitions to the Online Experiments. In this phase, the Online Agent deploys the proposed parameters into the production environment and evaluate them based on the large-scale Online Experiments.

\subsubsection{Online Experiment Code Generation}
To bridge the gap between abstract parameter values and production systems, the Online Agent uses the approved configuration candidates from the reasoning loop. It then automatically 
use these proposed parameters to generate the specific executable code, scripts, or configuration files required by the  recommender system's infrastructure following the restrictions in the skill.

\subsubsection{A/B Testing Task Generation}
Following code generation, the Online Agent automatically schedules the new online experiment on the production A/B testing platform. This step involves configuring the experiment's logistical parameters, such as allocating the appropriate percentage of user traffic, defining the control group (the baseline configuration) versus the treatment groups (the agent's proposed configurations), and setting the required time horizon for the experiment to reach statistical significance. Before online A/B testing starts, the users' review is required.

\subsubsection{A/B Testing Results Collection}
Upon the conclusion of the online Experiment, the Online Agent calls the platform API of online experiment to gather the final performance metrics (e.g., the North Star metrics and statistical significance). Crucially, this empirical performance data is then written back into the Agent Memory by updating the corresponding JSON files. This step enriches the agent's historical context and providing concrete ground-truth data to inform the next cycle of prompt construction and skillhub self-improvement.

\subsection{Agent Memory} \label{memory_update}
Agent Memory is the mechanism that allows multiple agents to retain and recall information across interactions. We design an agent memory module to serve as the central knowledge repository throughout the optimization life-cycle. 

\subsubsection{Memory Write and Read}
As illustrated in the workflow Figure \ref{Workflow}, the memory acts as a critical bridge between the reasoning loop and the online execution environment. It records the final, refined configuration candidates generated by the Critic Agent (via the Write operation). Each task includes id name, config string, explanation, proposed time, status, results, and evaluation check info. After online experiment, the Online Agent will update the Agent Memory and add results to the corresponding task item. In addition, elite task item will be read by the actor agent in the prompt generation in the next iteration of configuration optimizations.  

\subsubsection{Memory Pruning and Selection}
To optimize memory capacity, the Insight Agent periodically prunes redundant activity logs. Candidates are continuously evaluated and ranked based on their online experimental performance and categorical diversity. The system maintains a "top performers" pool by systematically discarding candidates that are strictly outperformed by others. This continuous filtering yields a refined subset where no single candidate definitively outperforms another across all metrics. By archiving only these elite candidates, the Insight Agent ensures the memory remains lean yet high-performing.

\subsubsection{Diversity Maximization}
To prevent redundancy, the Insight Agent maximizes the distance between candidates. The system first standardizes all results of each candidate, ensuring metrics with larger numerical scales do not disproportionately influence the selection. Using a greedy selection strategy, the algorithm identifies a strong starting point and iteratively selects subsequent items that are mathematically furthest from the existing set. 

\subsubsection{Pattern Learning Mechanisms}
The Insight Agent refines its historical context through two primary mechanisms:
\begin{itemize}
\item \textbf{Self-Learning}: 
The Insight Agent persistently interact with logs interactions, reasoning traces, and outcomes. Then agent searches common successful mods and configuration difference. 
Finally, the Insight Agent detects sensitive parameters with the highest impact across iterations and extracts patterns of primary metrics from individual tasks to inform future optimizations. For instance, if the memory indicates that aggressively increasing a certain diversity penalty consistently degrades overall user engagement, the Insight Agent extracts this insight as a new pattern.
\item \textbf{Cross-Learning}: The Insight Agent Utilize a MapReduce strategy, the agent learns patterns of primary metrics from multiple tasks in parallel (Map) followed by a global synthesis of patterns (Reduce). All involved files are listed in the memory.
\end{itemize}

This persistent state allows the agent to build a long-term understanding of the parameter space, preventing the repetition of failed configurations and enabling data-driven evolution across iterations.


\subsection{Self-Evolving Skillhub} \label{skillhub}
In the AgenticRecTune framework, each skill act as a plugin that integrate tools, knowledge, and workflows to enhance the optimization capabilities. We built an skillhub, where each skill is tailored to optimize a specific task, as shown in Figure. To ensure the AgenticRecTune fully comprehends the nuances of a given task, each skill includes the following components:

\begin{itemize}
\item \textbf{Task Context:} 
It  introduce the production or the recommender system context and specify which component in the recommender system we aim to optimize (e.g., pre-ranking stage configuration or system diversity parameters). It also provides the precise definition and operational impact of each parameter within the configuration.
\item \textbf{Task Requirement:} It defines the constraints and expected outputs for the optimization task, such as the allowed search space (value ranges) for tuning parameters, the required JSON schema for deploying the new configuration, and any infrastructure constraints the system must maintain. 
\item \textbf{North Star Metric:} It defines the primary optimization objective and secondary metrics, which help agent understand the priority of metrics and optimization direction (i.e, metric increase or decrease)
\item \textbf{Initial configuration parameters:} It supplies the current baseline or default configuration values actively deployed in the production environment. This serves as the anchor and starting point for the agent's exploration trajectory.
\item \textbf{Domain knowledge:} It injects task-specific heuristics, historical optimization logs, and expert guidelines. For instance, the agent is provided with the knowledge that excessively increasing a diversity penalty might degrade overall engagement, which helps narrow down the search space efficiently and prevents catastrophic configurations. Most importantly, the Skill Agent will automatically adds extra valuable learned knowledge.
\item \textbf{Tools:} It equips the agent with the executable functions necessary to complete the iterative optimization loop. This includes API tools to deploy new configurations to the A/B testing platform, query online experimentation results, and perform statistical significance analysis on the metrics.
\end{itemize}

While the static components of the Skill Hub (Section \ref{skillhub}) provide the initial foundational knowledge for the agent, the dynamic complexities of recommender systems necessitate continuous learning from the live experiment iterations. To achieve this, AgenticRecTune employs a self-evolving mechanism powered by the interplay between the Insight Agent (Section \ref{memory_update}) and the Skill Agent. As the Insight Agent systematically updates and summarizes empirical outcomes from historical iterations, the Skill Agent analyzes this centralized repository to autonomously refine its internal logic before deploying subsequent experiments.
This evolution operates through two core mechanisms:

\subsubsection{Dynamic Knowledge Extraction.}
Once the Insight Agent has analyzed historical logs and completed its self/cross-learning phases, it synthesizes learned patterns across related tasks. The Skill Agent then identifies patterns relevant to specific skills and automatically appends these refined rules to the Domain Knowledge component of that skill. Simultaneously, it dynamically tightens the search space bounds within the Task Requirements. And it ensures the reasoning loop avoids redundant exploration of previously identified suboptimal configurations.

\subsubsection{Novel Skill Generation.} 
Beyond merely appending insights, the Skill Agent leverages its accumulated memory and existing skill set to synthesize entirely new operational skills that guide the next optimization round. It mainly update the domain knowledge section. This allows the system to develop sophisticated tuning strategies without requiring human intervention to manually author new skill definitions. At the same time, engineers can explicitly track learned pattern.

With this self-evolving ability by Insignt Agent and Skill Agent, the Skillhub moves beyond static instructions to become a learning engine. This ensures that new proposed configurations are guided by past experience before they are tested in new real-world experiments.

\begin{table*}[!ht]
  \caption{Online A/B Testing Results: Task vs. Stages vs. Topline Metrics. Different Tasks in pre-ranking, ranking and re-ranking are considered.
  AgenticRecTune demonstrates a significant increase in all production topline metrics}
  \label{tab:stage}
  \begin{tabular}{ccccc}
    \toprule
    Task & Stage & Engagement Metric 1 & Engagement Metric 2 & Diversity Metric\\
    \midrule
    Value-Based Retrieval  & Pre-Ranking & 0.75\% & 0.90\% & 0.48\%\\
    Value Fusion & Ranking & 0.62\% & 0.19\% & 0.06\%\\
    Diversity & Re-Ranking & 0.21\% & 0.29\% & 3.43\% \\
  \bottomrule
\end{tabular}
\end{table*}

\begin{table*}[h]
  \caption{Online A/B Test Results: Agent Models vs. Topline Metrics. Pro models (Gemini 3 Pro and Gemini 1.5 Pro) have better performance compared with Flash model under the Diversity task in the re-ranking stage.}
  \label{tab:model}
  \begin{tabular}{cccc}
    \toprule
    Model Name & Engagement Metric 1 & Engagement Metric 2 & Diversity Metric\\
    \midrule
    Gemini 3 Pro   & 0.21\% & 0.29\% & 3.43\% \\
    Gemini 3 Flash & 0.08\%  & 0.07\% & 1.69\%\\
    Gemini 1.5 Pro & 0.22\% & 0.27\% & 2.11\%\\
  \bottomrule
\end{tabular}
\end{table*}

\begin{table*}
  \caption{Online A/B Test Results: Agent Strategies vs. Topline Metrics. Actor-Critic Strategy demonstrates superior efficacy in Value-Based Retrieval tasks compared to singe agent strategy. }
  \label{tab:strategy}
  \begin{tabular}{cccc}
    \toprule
    Model Name & Engagement Metric 1 & Engagement Metric 2 & Diversity Metric\\
    \midrule
    Actor-Critic Strategy  & 0.75\% & 0.90\% & 0.48\%\\
    Single Agent Strategy &0.29\% & 0.26\% & 0.06\%\\
  \bottomrule
\end{tabular}
\end{table*}

\section{Online Experiment}

\subsection{Experiment Setup}
We conducted online A/B testing to evaluate the proposed configurations within the production environment, including the pre-ranking, ranking, and re-ranking stages. Live user traffic was randomly partitioned into orthogonal buckets. In each round of online A/B testing, the control group is the existing tuned configurations on the production, for the current production, and treatment groups exposed to the parameter settings generated by AgenticRecTune. Each experiment was run as a standard launch period to ensure statistical significance (evaluated at a p-value < 0.05).

\subsection{Experiment Results}
The deployment of AgenticRecTune consistently yielded positive, statistically significant improvements against the topline metrics across three major stages.
\subsubsection{Pre-ranking Stage}
In the pre-ranking stage, the primary objective is to efficiently prune a large pool of candidates down to a smaller, high-quality subset using lightweight, multi-head machine learning models. These models assign multiple scores to each candidate to determine its importance. In value-based retrieval task, these distinct scores are combined via complex weighted strategy to provide a final score.

In the pre-ranking stage, we apply AgenticRecTune to optimize the configurations in the value-based retrieval task, which merges the scores from pre-ranking model predictions. By loading the task context, engineering experience and tools in the corresponding skill, the AgenticRecTune starts from the production configuration and propose different configurations as the treatment group.
AgenticRecTune dynamically prioritize adjusting sensitive and primary weights while allocating less treatment arms to insensitive score. After rounds of small traffic of iteration, AgenticRecTune succesfully search the better candidate compared with baseline. Based on the Table \ref{tab:stage}, AgenticRecTune yield a statistically significant increase in the production engagement metric 1 \& 2 and diversity metric in the Value-Based Retrieval task. 

\subsubsection{Ranking Stage}
The ranking stage is computationally intensive and presents a highly sensitive, multi-objective optimization problem. The recommender system must balance competing goals, ranging from short-term user engagement metrics to long-term objectives. During the ranking model's score fusion stage, configurations are used to balance scores across different prediction targets. Each candidate is comprehensively evaluated against multiple criteria, such as quality and relevance. The value fusion task applies a specific configuration to compute a weighted combination of these scores.

Human tuning and data-driven learning rely on extensive domain experience to discover optimal fusion weights for these distinct sub-models. In contrast, AgenticRecTune leverages its reasoning loop and vast amounts of historical data to explore this high-dimensional weight space significantly more efficiently than traditional grid search methods. The agent can  synthesizes optimization strategies based on top-performing historical configurations and production development requirements, achieving an improvement that boosts production topline metrics. As demonstrated in Table \ref{tab:stage}, AgenticRecTune successfully identifies superior parameters and improve the engagement Metric 1 \& 2. 

\subsubsection{Re-ranking Stage}
The focus of re-ranking stage shift towards optimizing list-wise properties, including topic diversity, business logic, and the mitigation of content fatigue. We use the diversity task as an example, which integrates multiple strategies and requires the precise calibration of numerous thresholds and weights.

Through iterative feedback loops, AgenticRecTune discovered the better configuration that successfully diversified the user feed. 
Table \ref{tab:stage} shows that diversity metric increases by 3.43\% and engagement Metric 1 \& 2 also increase.   
Notably, it achieved this improvement in long-term ecosystem health without triggering the short-term engagement drops that frequently plague aggressive diversity adjustments. Furthermore, given the high dimensionality of the parameter space, human engineers can easily overlook the importance of specific variables during manual tuning. We observed that AgenticRecTune provides novel insights into configuration tuning, identifying impactful parameter interactions that engineers had previously missed.

\subsection{Model Ablation Study}

To isolate and understand the impact of the underlying LLM's reasoning capabilities, we conducted an ablation study comparing three different foundation models driving the agent: Gemini 3 Pro, Gemini 3 Flash, and Gemini 1.5 Pro.

We use the diversity task in the pre-ranking stage. The results, summarized in Table \ref{tab:model}, indicate that model scale and architectural generation significantly influence agent performance. Gemini 3 Pro demonstrated the most balanced performance across all dimensions, particularly excelling in diversity metric with a significant 3.43\% improvement. While Gemini 1.5 Pro showed competitive results in the first two metrics, its lower performance in diversity metric (2.11\%) suggests that the newer Gemini 3 architecture is better suited for the complex reasoning required in this optimization task. Conversely, the Gemini 3 Flash model, while computationally efficient, showed the lowest gains across all metrics, confirming that high-parameter reasoning capabilities are essential for the agent to navigate the search space effectively.

\subsection{Agent Strategies Ablation Study}

To improve the robustness of our framework and allow the agent to explore more diverse conditions efficiently, we propose an Actor-Critic Agent strategy. In this configuration, the Actor agent first proposes multiple candidates and provides a detailed explanation for each. Subsequently, a Critic agent evaluates these proposals, providing constructive feedback and filtering out sub-optimal solutions before they are executed. We verify the actor-critic strategy under the Value-Based Retrieval task.

Table \ref{tab:strategy} shows that the actor-critic strategy outperforms the baseline single-agent strategy. The Actor-Critic configuration achieved a 0.75\% and 0.90\% lift in engagement metric 1 and 2, respectively, which is more than double than the gains observed with the Actor-only agent. This suggests that the iterative feedback loop provided by the Critic agent effectively mitigates "hallucinations" in the optimization process and encourages more rigorous candidate selection. Interestingly, while the Actor-Critic model improved engagement metric 1 and 2 substantially, diversity metric saw a more modest improvement (0.48\%), indicating that the primary benefit of the dual-agent architecture lies in its ability to refine precision rather than purely expanding the exploration breadth.

\section{Conclusion}


In this paper, we presented AgenticRecTune, a multi-agent framework designed to navigate the industrial-scale recommender system configurations tuning. This work extends beyond individual model enhancements to span multiple stages, including pre-ranking, ranking, and re-ranking. Traditional methods, ranging from manual tuning and grid search to data-driven learning, rely heavily on labor-intensive processes and deep domain expertise to identify optimal fusion weights. By orchestrating a specialized collective of Actor, Critic, Insight, Skill, and Online agents, we have successfully automated the end-to-end configuration optimization workflow, effectively bypassing the scalability bottlenecks of traditional human-driven iterations.

Our framework introduces a self-evolving SkillHub that transforms raw experimental data into executable domain expertise. Through the closed-loop synergy between the Insight and Skill agents, the system continuously refines its search space and synthesizes novel operational strategies from historical outcomes. Furthermore, by employing an Actor-Critic architecture and an Online Agent to target real-world Pareto fronts, AgenticRecTune ensures high-precision exploration that aligns with multi-objective product targets.
The practical utility of this system is underscored by its successful deployment on Google Discover. As recommender systems continue to grow in complexity, AgenticRecTune provides a self-improving framework for the configurations optimization in industrial-scale environments with multiple shifting target metrics.


\bibliographystyle{ACM-Reference-Format}
\bibliography{sample-base}

\newpage

\appendix











\end{document}